\documentclass{elsart}

 \usepackage{graphicx}

\usepackage{amssymb}

\begin{document}

\begin{frontmatter}



\title{Two Models for Bolometer and Microcalorimeter Detectors with 
Complex Thermal Architectures}

\author{J. W. Appel}
\author{M. Galeazzi}

\address{ University of Miami, Department of Physics, P.O. Box 248046, 
Coral Gables, FL 33124 USA}

\date{\today}

\begin{abstract}

We have developed two analytical models to describe the performance of 
cryogenic microcalorimeters and bolometers. One of the models is suitable 
to describe Transition Edge Sensor(TES) detectors with an integrated absorber, 
the other is suitable for detectors with large absorbers. Both models take 
into account hot-electron decoupling and absorber decoupling. The differential 
equations describing these models have been solved using block diagram algebra.
 Each model has produced closed form solutions for the detector's responsivity,
 dynamic impedance, and noise equivalent power for phonon noise, Johnson noise,
 amplifier noise, 1/f noise, and load resistor noise.
\end{abstract}
\end{frontmatter}

\section{Introduction}

The operational principle of simple microcalorimeters and bolometers is 
based on three components. An absorber where the incoming power or energy 
is dissipated and converted into a change in temperature, a sensor that 
reads the change in temperature, and a thermal link from the detector 
to the heat sink that brings the system back to equilibrium after a 
measurement. The sensor is usually a resistor whose resistance depends 
strongly on temperature around the working point. In this case a change 
in resistance can be measured as a change in voltage or a change in current 
using a current or voltage bias.

In 1982 J.C Mather \cite{mather1} presented a complete non-equilibrium theory 
for the noise in simple bolometers with ideal resistive thermometers and 
in 1984 it was extended to microcalorimeter performance \cite{mather2}. At 
temperatures below ~200 mK the ideal assumptions are no longer valid and 
complex thermal architectures are needed to understand the behavior of these 
devices. At these low temperatures the thermal fluctuations between the 
thermometer lattice and its electron system, or between the thermometer and 
the absorber are no longer negligible and therefore these components must 
be considered as separate entities in the model. These non-ideal effects 
are called electron decoupling and absorber decoupling. Another consequence 
of working at low temperatures is the increased dependence of the thermometer 
resistance on the readout power, making the ideal resistance-temperature 
relationship inaccurate. Galeazzi and McCammon \cite{GM} constructed a general 
procedure for developing bolometer and microcalorimeter models for these 
complex thermal architectures using block diagram formalism of control theory.

 To quantify the relation between incoming power or energy and the measured 
change in voltage or current, including non-ideal effects, this paper follows 
the modeling procedure of Galeazzi and McCammon \cite{GM}.  The first step in 
this modeling procedure is to set up the temperature equations, then apply a 
Taylor expansion to derive a linear model for small temperature deviations 
from equilibrium. Afterward, Fourier transforms are used to express the 
equations in the frequency domain, and finally the coupled equations are 
solved using block diagram algebra. This procedure yields closed form 
solutions for the responsivity and dynamic impedance of the model, including 
noise contributions.

The two models developed in this paper are modifications of the absorber 
decoupling model obtained in \cite{GM}. Model 1 was developed to describe 
the new generation of transition edge sensor detectors where the absorber 
is not electrically isolated from the thermometer by a gluing agent but 
rather the two are deposited one on top of the other \cite{M1}. Model 2 
describes microcalorimeters that have the heat sink connected to the absorber 
instead of to the thermometer. This may occur when the absorber is much bigger 
than the thermometer and therefore it is necessary to connect the heat 
sink to the absorber rather than to the thermometer. These two models will 
help optimize the next generation of detectors, and because of the analytical 
results of the modeling procedure the relations between the detector's 
resolution and the different parameters included in the model should be clear.

\section{Model 1}

\begin{figure}
\begin{center}
\includegraphics[width=0.3\textwidth]{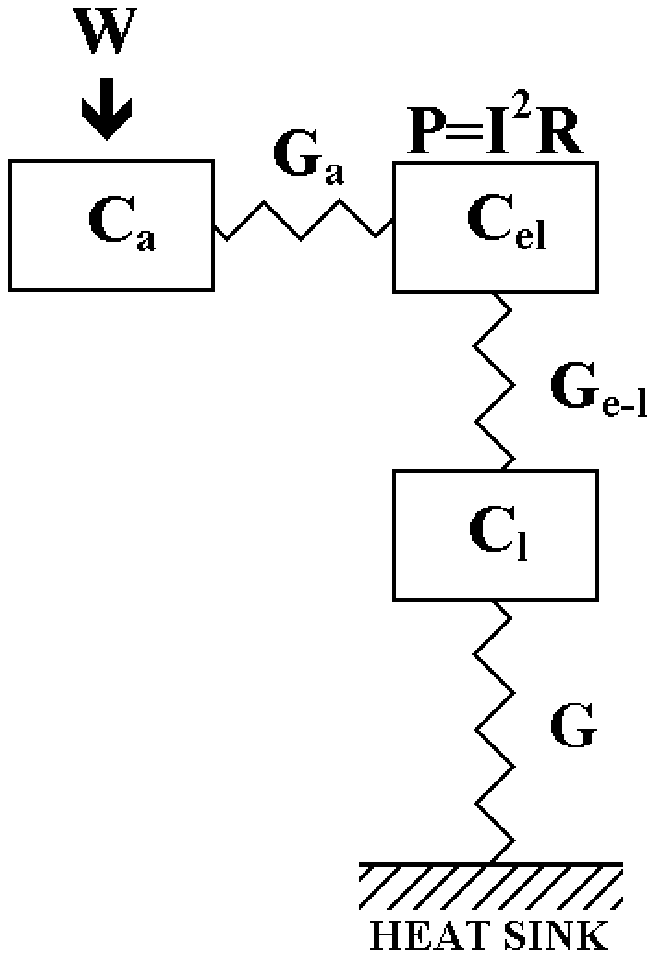}
\end{center}
\caption{\footnotesize Thermal architecture of Model 1.}
\label{TA1}
\end{figure}

This model is suitable to describe TES detectors with an integrated absorber. 
In this model both the electron system and the absorber are detached from 
the lattice. The lattice is connected to the heat sink by a thermal 
conductivity $G$, the electron system is connected to the lattice by a thermal 
conductivity $G_{e-l}$, and the absorber is connected to the electron system 
by a thermal conductivity $G_{a}$ ( see Fig.~\ref{TA1}). The absorber is 
directly connected to the electron system instead of the lattice because 
with integrated absorbers the absorber-lattice thermal coupling is expected 
to be negligible compared to that of the absorber-electron system.

\subsection{Responsivity S($\omega$)}

The following equations determine the temperature for each of the three 
components in the model:
\begin{equation}
C_{a}\frac {d(T'_{a})}{dt}+\int_{T'_{e}}^{T'_{a}} G_{a}(T')dT'=W
\label{eq1}
\end{equation}
\begin{equation}
C_{e}\frac {d(T'_{e})}{dt}+\int_{T'_{a}}^{T'_{e}} G_{a}(T')dT'+
\int_{T'_{l}}^{T'_{e}} G_{e-l}(T')dT'=P(T'_{e})
\end{equation}
\begin{equation}
C_{l}\frac {d(T'_{l})}{dt}+\int_{T'_{e}}^{T'_{l}} G_{e-l}(T')dT'+
\int_{T_{s}}^{T'_{l}} G(T')dT'=0,
\end{equation}
where $C_{a}$, $C_{e}$, and $C_{l}$ are the heat capacities of the absorber, 
the electron system, and the lattice system respectively, and $T'_{a}$, 
$T'_{e}$, and $T'_{l}$ are the corresponding temperatures. $W$ is the incoming 
outside power to be measured and $P(T'_{e})$ is the Joule power dissipated 
into the sensor by the bias current/voltage. In the case of microcalorimeters 
$W=E \delta (t_{o})$, where $E$ is the photon energy and $\delta(t_{o})$ is the
 delta function.

The equilibrium conditions of the system are obtained by setting the outside 
power to zero ($W=0$), and $d(T'_{x})/dt=0$ ($x$= $a$, $e$, or $l$) since the 
equilibrium temperatures are independent of time. Therefore the equilibrium 
temperatures $T_{a}$ of the absorber, $T_{e}$ of the electron system, and 
$T_{l}$ of the lattice are given by the integrals in the previous three 
equations. For example the integral in Eq.~\ref{eq1} must equal zero at 
equilibrium, which implies that the thermal equilibrium temperature of the 
absorber is the same as that of the electron system. We are interested in 
small deviations about the equilibrium temperatures, therefore we set 
$T'_{x}=T_{x}+\Delta T_{x}$, where $T_{x}$ is the equilibrium temperature 
for each component of the model, and $\Delta T_{x}$ is the small temperature 
deviation from equilibrium:
\begin{equation}
C_{a}\frac {d(T_{a}+\Delta T_{a})}{dt}+\int_{T_{e}+\Delta T_{e}}^{T_{a}+
\Delta T_{a}} G_{a}(T')dT'=W
\end{equation}
\begin{equation}
C_{e}\frac {d(T_{e}+\Delta T_{e})}{dt}+\int_{T_{a}+\Delta T_{a}}^{T_{e}+
\Delta T_{e}} G_{a}(T')dT'
+\int_{T_{l}+\Delta T_{l}}^{T_{e}+\Delta T_{e}} G_{e-l}(T')dT'=P(T_{e}+
\Delta T_{e})
\end{equation}
\begin{equation}
C_{l}\frac {d(T_{l}+\Delta T_{l})}{dt}+\int_{T_{e}+\Delta T_{e}}^{T_{l}+
\Delta T_{l}} G_{e-l}(T')dT'
+\int_{T_{s}}^{T_{l}+\Delta T_{l}} G(T')dT'=0.
\end{equation}

In the small signal limit $\Delta T_{x}$ is small compared to $T_{x}$, and a 
Taylor expansion up to the first $\Delta T_{x}$ term is appropriate. The 
results are the equations that determine small temperature deviations about 
equilibrium:
\begin{equation}
 C_{a}\frac {d(\Delta{T_{a}})}{dt}+G_{a}\Delta T_{a}=W+G_{a}\Delta T_{e}
\label{de1}
\end{equation}
\begin{equation}
 C_{e}\frac {d(\Delta{T_{e}})}{dt}+G_{a}\Delta T_{e}+ G_{e-l}(T_{e})
\Delta T_{e}=\Delta P
+ G_{a}\Delta T_{a}+ G_{e-l}(T_{l})\Delta T_{l}
\label{de2}
\end{equation}
\begin{equation}
C_{l}\frac {d(\Delta T_{l})}{dt}+G_{e-l}(T_{l})\Delta T_{l} + G\Delta T_{l}=
G_{e-l}(T_{e})\Delta T_{e},
\label{de3}
\end{equation}
where $\Delta P=P(T_{e}+\Delta T_{e})-P(T_{e})$, and for simplicity we 
used $G_{a}=G_{a}(T_{a})$ and $G=G(T_{l})$.

These are coupled differential equations which are difficult to solve directly;
 instead they are transformed into coupled algebraic equations using Fourier 
transforms. The quantity $\Delta P$ represents what is known as the 
electro-thermal feedback term and can be written as $\Delta P=-G_{ETF} 
\Delta T_{e}$, where $G_{ETF}=P(R-R_{L})\alpha/T_{e}R(R_{L}+R)$; (see 
reference \cite{GM}). Converting Eqs.~\ref{de1}, \ref{de2}, and \ref{de3} 
into the frequency domain using Fourier transforms we obtain:
\begin{equation}
 j\omega C_{a}\Delta T_{a}+G_{a}\Delta T_{a}=W+G_{a}\Delta T_{e}
\label{FD1}
\end{equation}
\begin{equation}
j\omega C_{e}\Delta T_{e}+(G_{a}+ G_{e-l}(T_{e})+G_{ETF})\Delta T_{e}=
G_{a}\Delta T_{a}
+ G_{e-l}(T_{l})\Delta T_{l}
\label{FD2}
\end{equation}
\begin{equation}
j\omega C_{l}\Delta T_{l}+(G_{e-l}(T_{l}) + G)\Delta T_{l}=G_{e-l}(T_{e})
\Delta T_{e}.
\label{FD3}
\end{equation}

\begin{figure}
\begin{center}
\includegraphics[width=0.2\textwidth]{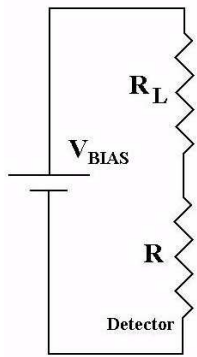}
\end{center}
\caption{\footnotesize Typical readout circuit. Notice that if $R_{L}<<R$ 
the detector is voltage biased, 
if $R_{L}>>R$ the detector is current biased. }
\label{figcircuit}
\end{figure}

With these equations it is possible to solve for $\Delta T_{e}$, which can be 
related to the measured quantities $\Delta I$ or $\Delta V$, using the typical 
detector readout circuit of Fig.~\ref{figcircuit}:

\begin{equation}
\Delta V=V\frac{\alpha}{T_{e}} \frac{ R_{L}}{R_{L}+R}\Delta T_{e}
\label{c1}
\end{equation}
\begin{equation}
\Delta I=-I\frac{\alpha}{T_{e}} \frac{ R}{R_{L}+R}\Delta T_{e},
\label{c2}
\end{equation}
where $\alpha=T_{e}/R \times dR/dT_{e}$ is the sensitivity of the detector, 
$R_{L}$ is the load resistance, $R$ is the resistance of the detector, $V$ is 
the voltage across $R$, and $I$ is the current flowing through $R$.

\begin{figure}
\begin{center}
\includegraphics[width=1\textwidth]{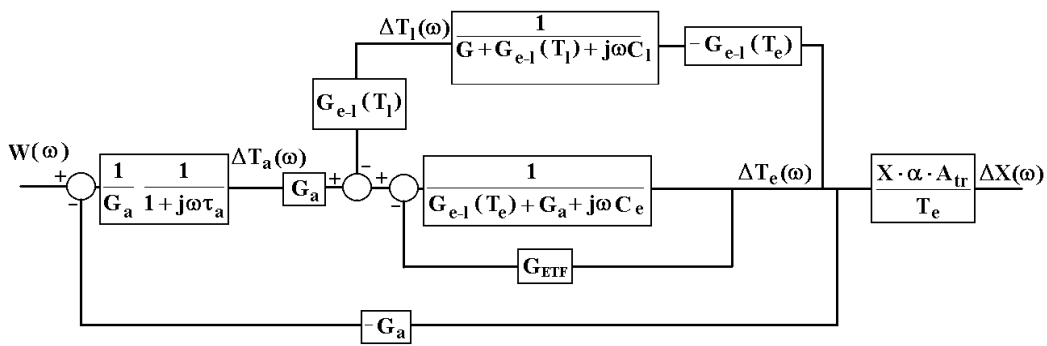}
\end{center}
\caption{\footnotesize Block diagram representing Model 1.}
\label{BD1}
\end{figure}

To simplify the notation let $X$ be either $V$ or $I$, and introduce the 
quantity $A_{tr}=R/X \times dX/dR$ to be deduced from the  previous two 
equations. Then Eqs. \ref{c1} and \ref{c2} can be summarized as:
\begin{equation}
\frac{\Delta X}{X}=\alpha A_{tr}\frac{\Delta T_{e}}{T_{e}}.
\label{c3}
\end{equation} 
Equations \ref{FD1}, \ref{FD2}, \ref{FD3}, and \ref{c3} can be solved using 
the block diagram of Fig.~\ref{BD1}. To set up the block diagram consider the 
left hand side of Eqs.\ref{FD1}, \ref{FD2}, \ref{FD3}, and \ref{c3} as the 
response function of the absorber system, electron system, lattice system, and 
circuit readout respectively. The right hand side of these equations 
corresponds to the input to each system. Connecting the response functions 
with their appropriate inputs leads to the block diagram of Fig.~\ref{BD1}.

To solve the block diagram in Fig.~\ref{BD1} for $\Delta X(\omega)$ we used 
the procedure and simplification rules of the block diagram formalism described
 in \cite{GM}. This result is then used to find the responsivity, which is 
defined as $S(\omega)=\Delta X(\omega)/W(\omega)$. The following responsivity 
characterizes the response of Model 1 detectors:
\begin{equation}
S(w)=\frac{G_{a}}{(G_{a}+j\omega C_{a})
\bigg[(G_{a}+G_{e-l}(T_{e})+G_{ETF}+j\omega C_{e})-
\frac{G_{e-l}(T_{l})G_{e-l}(T_{e})}{j\omega C+G_{e-l}(T_{l})+G}\bigg]
-G_{a}^{2}}\frac{X\alpha A_{tr}}{T_{e}}.
\label{Sw}
\end{equation}

\subsection{Dynamic Impedance}
A detector can also be described by its complex dynamic impedance 
$Z(\omega)=dV(\omega)/dI(\omega)$. The dynamic impedance differs from the 
detector resistance due to the effect of the electro-thermal feedback. When 
the current changes, the power dissipated into the detector changes too, 
therefore the temperature and the detector's resistance change. The dynamic 
impedance is a useful parameter because it is easily measured experimentally. 
To find the dynamic impedance we use $-G_{ETF}\Delta T_{e}=\Delta P$ in 
Eqs.~\ref{FD2}, and use Eqs.~\ref{FD1}, \ref{FD2}, and \ref{FD3} to find 
$\Delta T_{e}$ in terms of $\Delta P$, $\omega$, the heat capacity of each of 
the three components, and the three thermal conductivities:
\begin{equation}
dT_{e}=\frac {dP}{\bigg(j\omega C_{e}+G_{a}+G_{e-l}(T_{e})-\frac {G_{a}^{2}}
{G_{a}+j\omega C_{a}}-\frac{G_{e-l}(T_{l})G_{e-l}
(T_{e})}{j\omega C+G_{e-l}(T_{l})+G}\bigg)}.
\label{DI1}
\end{equation}
Differentiating Ohm's law ($V=IR$) and using the definition of sensitivity 
$\alpha$ we obtain: 
\begin{equation}
dV=RdI+I\frac {\alpha RdT}{T}.
\label{DI2}
\end{equation}
Substituting Eq.~\ref{DI1} into Eq~\ref{DI2} and using the fact that 
$dP=VdI+IdV$, it is possible to solve for $dV/dI$ and obtain the following 
result for the dynamic impedance: 
\begin{equation}
Z(\omega)=\frac {dV}{dI}=R\frac{\bigg[\alpha P+T\bigg(j\omega C_{e}+G_{a}+
G_{e-l}(T_{e})-\frac {G_{a}^{2}}{G_{a}+j\omega C_{a}}-\frac{G_{e-l}(T_{l})
G_{e-l}(T_{e})}{j\omega C+G_{e-l}(T_{l})+G}\bigg)\bigg]}{\bigg[-\alpha P+
T\bigg(j\omega C_{e}+G_{a}+G_{e-l}(T_{e})-\frac {G_{a}^{2}}
{G_{a}+j\omega C_{a}}-\frac{G_{e-l}(T_{l})G_{e-l}(T_{e})}{j\omega C+G_{e-l}
(T_{l})+G}\bigg)\bigg]}.
\end{equation}
\subsection{Noise}
The effect of noise on a detector's performance is quantified by the Noise 
Equivalent Power (NEP). It corresponds to the power that would be required as 
input to the detector in order to generate an output equal to the signal 
generated by the noise. The NEP can therefore be calculated as the ratio 
between the output generated by the noise and the responsivity of the detector:
\begin{equation}
 NEP_{y}=\frac {\Delta X_{y}}{S(\omega)}.
\end{equation} 
The variable $y$ stands in for any of the possible noise terms: $amp$=amplifier
 noise, $j$=Johnson noise, $R_{L}$=load resistor noise, $1/f$=$1/f$ noise, 
$a$=absorber-electron system thermal noise, $th$=heat sink-lattice thermal 
noise or $he$=electron system-lattice thermal noise. 
 
\begin{figure}
\begin{center}
\includegraphics[width=1\textwidth]{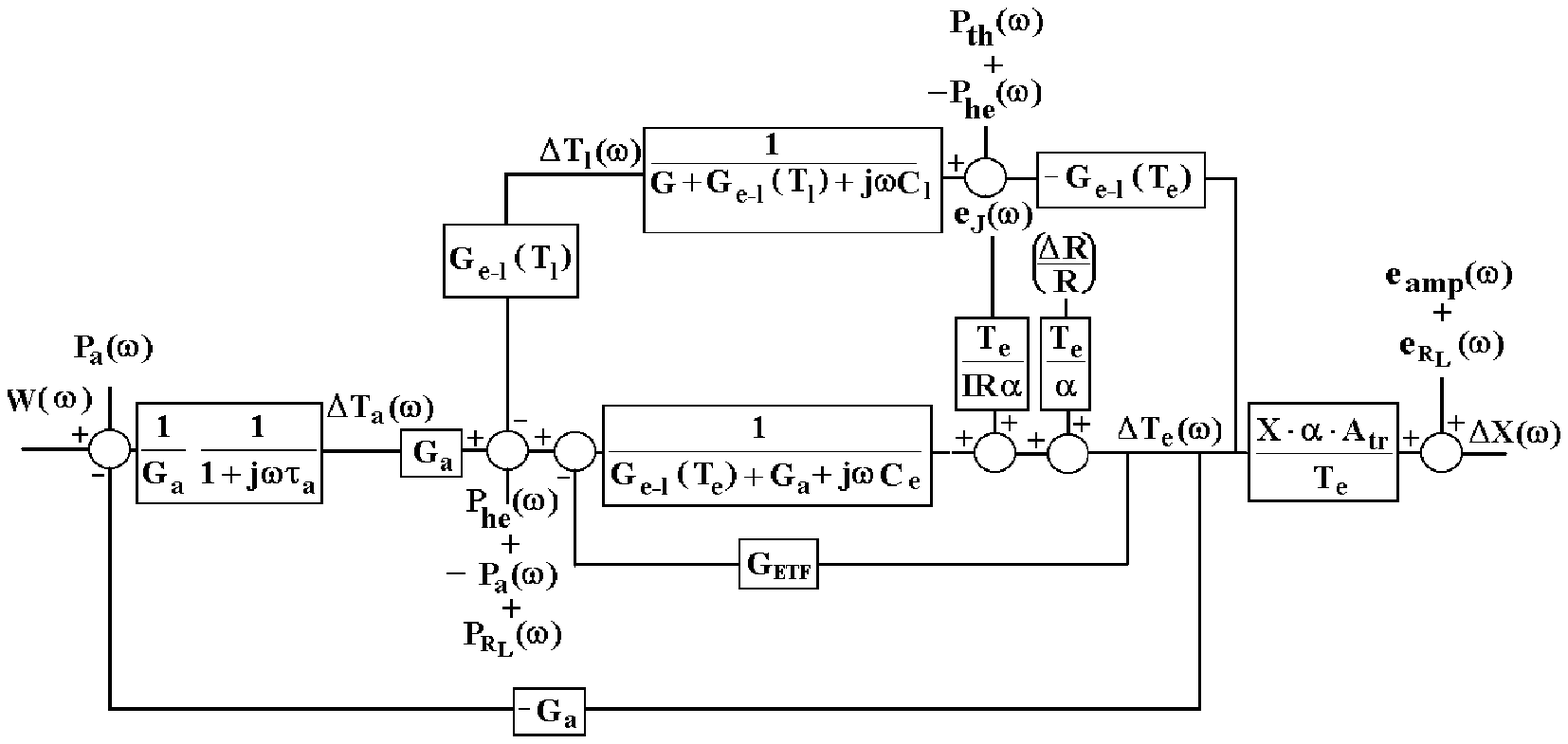}
\end{center}
\caption{\footnotesize Block diagram including noise contributions 
for Model 1.}
\label{BD2}
\end{figure}

To obtain the Noise Equivalent Power for each term, the noise contributions 
$e_{amp}$, $e_{j}$, $e_{R_{L}}$, $P_{R_{L}}$, $(\Delta R/R)_{1/f}$, $P_{a}$, 
$P_{th}$, and $P_{he}$ must be added to the block diagram of Fig.~\ref{BD1}.
 Figure~\ref{BD2} shows where each noise term should be added to the block
 diagram (for more details see~\cite{GM}). Solving the noise block diagram for 
each noise term independently and dividing by the responsivity obtained 
in Eq.~\ref{Sw} we obtain the following NEP's: 
\begin{equation}
NEP_{a}=P_{a}(\omega)j\omega\tau_{a}
\end{equation}
\begin{equation}
NEP_{R_{L}}=P_{R_{L}}(\omega)(1+j\omega \tau_{a})+\frac {e_{R_{L}}}{S(\omega)}
\end{equation}
\begin{equation}
NEP_{amp}=\frac {e_{amp}}{S(\omega)}
\end{equation}
\begin{equation}
NEP_{he}=P_{he}\frac{(1+j\omega \tau_{a}) (G+j\omega C_{l})}{(G+G_{e-l}(T_{l})
+j\omega C_{l})}
\end{equation}
\begin{equation}
NEP_{th}=P_{th}\frac {G_{e-l}(T_{l})(1+j\omega \tau_{a})}{(G+G_{e-l}(T_{l})
+j\omega C_{l})}
\end{equation}
\begin{eqnarray}
NEP_{e_{j}}=e_{j}(\omega)\frac {T_{e}}{IR\alpha}(1+j\omega \tau_{a}) 
\bigg(G_{a}+G_{e-l}(T_{e})+j\omega C_{e}\nonumber-\\ 
\frac{G_{e-l}(T_{l})G_{e-l}(T_{e})}{j\omega C+G_{e-l}(T_{l})+G}-
\frac {G_{a}^2}{G_{a}+j\omega C_{a}}\bigg)
\end{eqnarray}
\begin{eqnarray}
NEP_{1/f}=\Big(\frac {\Delta R(\omega)}{R}\Big)_{1/f}\frac {T_{e}}{\alpha}
(1+j\omega\tau_{a}) \bigg(G_{a}+G_{e-l}(T_{e})+j\omega C_{e}\nonumber-\\ 
\frac{G_{e-l}(T_{l})G_{e-l}(T_{e})}{j\omega C+G_{e-l}(T_{l})+G}-\frac 
{G_{a}^2}{G_{a}+j\omega C_{a}}\bigg).
\end{eqnarray} 
Where $\tau_{a}=C_{a}/G_{a}$.

\section{Model 2}
\begin{figure}
\begin{center}
\includegraphics[width=0.3\textwidth]{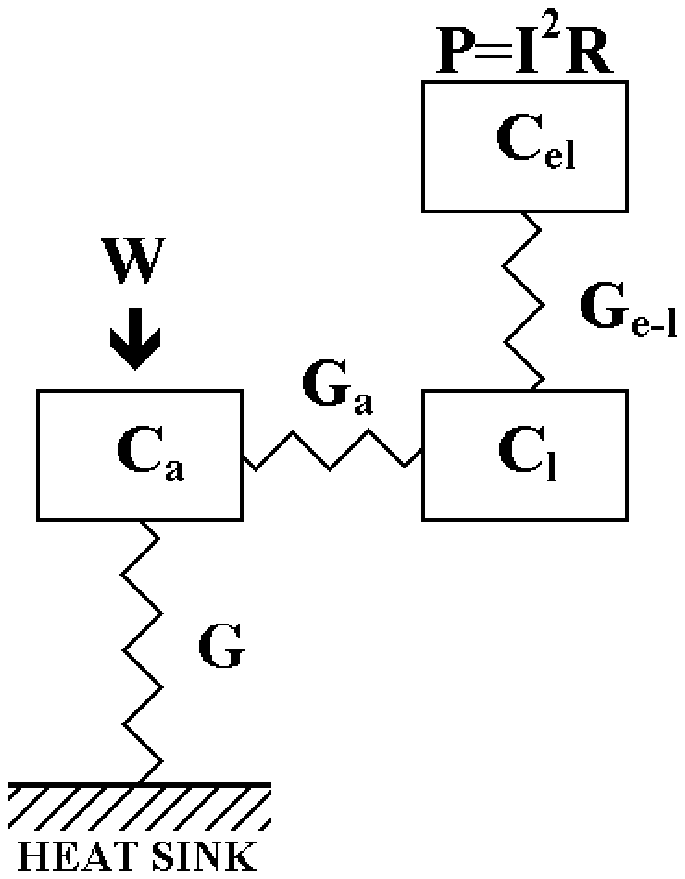}
\end{center}
\caption{\footnotesize Thermal architecture of Model 2.}
\label{TA2}
\end{figure}

In experiments involving dark matter detectors and double-beta decay detectors 
the absorber size is significant and can have a mass up to almost 1 Kg 
\cite{DM}. For mechanical reasons these large absorbers must be mechanically
 connected to the heat sink. The thermal link between the detector and the 
heat sink can therefore also be through the absorber rather than through the 
sensor. Model 2 reflects this condition by having the absorber connected to 
the heat sink through a thermal conductivity $G$. Model 2 also takes into 
account absorber decoupling and electron decoupling by connecting the lattice 
system to the absorber through a thermal conductivity $G_{a}$ and by having 
the electron system connected to the lattice system through a thermal 
conductivity $G_{e-l}$ (see Fig.~\ref{TA2}).

\begin{figure}
\begin{center}
\includegraphics[width=1\textwidth]{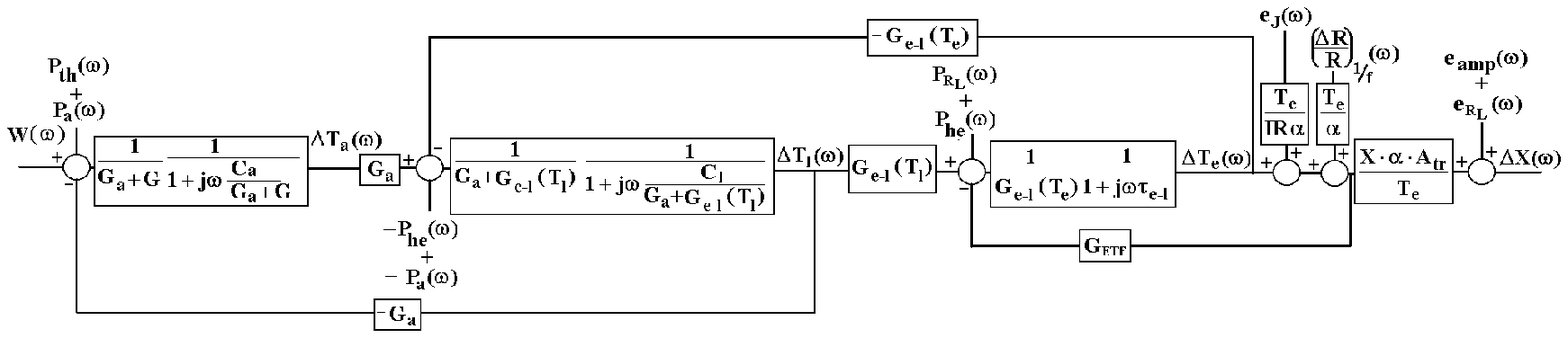}
\end{center}
\caption{\footnotesize Block diagram including noise contributions 
for Model 2.}
\label{BD3}
\end{figure}

Applying the same procedure previously used for Model 1 we obtain for Model 2 
the block diagram of Fig.~\ref{BD3}.Solving the block diagram for the 
responsivity, the dynamic impedance, and all the noise contributions, we 
obtain the following results:
\begin{equation}
\scriptstyle S(\omega)=\frac{G_{a}G_{e-l}(T_{l})}{(G_{e-l}(T_{e})+G_{ETF}+
j\omega C_{e})\bigg[(G_{a}+G_{e-l}(T_{l})+j\omega C_{l})(G_{a}+G
+j\omega C_{a})-G_{a}^{2}\bigg]-G_{e-l}(T_{l})G_{e-l}(T_{e})(G_{a}+G
+j\omega C_{a})}\frac {X\alpha A_{tr}}{T_{e}}
\end{equation}
\begin{equation}
Z(\omega)=R\frac{\alpha P+T_{e}\bigg[j\omega C_{e}+G_{e-l}(T_{e})-
G_{e-l}(T_{l})\bigg(\frac {G_{e-l}(T_{e})(j\omega C_{a}+G_{a}+G)}
{(j\omega C_{a}+G_{a}+G)(j\omega C_{l}+G_{a}+G_{e-l}(T_{l}))-G_{a}^2}
\bigg)\bigg]}{-\alpha P+T_{e}\bigg[j\omega C_{e}+G_{e-l}(T_{e})-G_{e-l}(T_{l})
\bigg(\frac {G_{e-l}(T_{e})(j\omega C_{a}+G_{a}+G)}{(j\omega C_{a}+G_{a}+G)
(j\omega C_{l}+G_{a}+G_{e-l}(T_{l}))-G_{a}^2}\bigg)\bigg]}
\end{equation}
\begin{equation}
NEP_{th}=P_{th}
\end{equation}
\begin{equation}
NEP_{amp}=\frac {e_{amp}}{S(\omega)}
\end{equation}
\begin{equation}
NEP_{a}=P_{a}\bigg[\frac {G+j\omega C_{a}}{G_{a}}\bigg]
\end{equation}
\begin{equation}
NEP_{he}=P_{he}\bigg[\frac {(G_{a}+G+j\omega C_{a})(G_{a}+G_{e-l}(T_{l})
+j\omega C_{l})-G_{a}^2}
{G_{a}G_{e-l}(T_{l})}-\frac {G+G_{a}+j\omega C_{a}}{G_{a}}\bigg]
\end{equation}
\begin{equation}
NEP_{R_{L}}=P_{R_{L}}\bigg[\frac {(G_{a}+G+j\omega C_{a})(G_{a}+G_{e-l}(T_{l})
+j\omega C_{l})-G_{a}^2}
{G_{a}G_{e-l}(T_{l})}\bigg]+\frac {e_{RL}}{S(\omega)}
\end{equation}
\begin{eqnarray}
NEP_{e_{j}}=e_{j}\frac {T_{e}}{IR\alpha} \frac {(G_{a}+G+j\omega C_{a})[(G_{a}
+G_{e-l}(T_{l})+j\omega C_{l})-\frac {G_{e-l}(T_{l})G_{e-l}(T_{e})}
{j\omega C_{e}+G_{e-l}(T_{e})}]-G_{a}^2}{G_{a}G_{e-l}(T_{l})}\nonumber\\ 
\times(j\omega C_{e}+G_{e-l})
\end{eqnarray}
\begin{eqnarray}
NEP_{\frac{1}{f}}=\Big(\frac {\Delta R(\omega)}{R}\Big)_{\frac{1}{f}} 
\frac {(G_{a}+G+j\omega C_{a})[(G_{a}+G_{e-l}(T_{l})+j\omega C_{l})-
\frac{G_{e-l}(T_{l})G_{e-l}(T_{e})}{j\omega C_{e}+G_{e-l}(T_{e})}]-
G_{a}^2}{G_{a}G_{e-l}(T_{l})}\nonumber\\ \times \frac {T_{e}}{\alpha}
(j\omega C_{e}+G_{e-l}).
 \end{eqnarray}
\section{Examples of Energy Resolution and Time constant Results from the 
Models}
In experimental setups of detectors some of the parameters are fixed while
 others can vary. With the freedom to vary a few parameters the goal is to 
optimize the detectors characteristics. The equations derived in this paper
offer the flexibility and power to perform such operations. An example of such
 applications is reported in fig. \ref{fig7}. The characteristics of interest
 are energy resolution and the time constant of the detector. 
The energy resolution is calculated following equation 69 in 
reference \cite{GM}. The time constant is calculated as the inverse 
of the first turn frequency of the responsivity. The variable 
parameters in the examples of fig. \ref{fig7} are: heat capacity of the 
absorber ($C_{a}$) vs. thermal conductivity 
between absorber and thermometer ($G_{a}$)
 and thermal conductivity between detector and heat sink ($G$) vs. thermal 
conductivity between absorber and thermometer ($G_{a}$). The fixed parameters
 are reported in Table 1. By inputing the fixed parameter into the two models
 of this paper and the model found in \cite{GM} and letting $C_{a}$, $G$ , 
and $G_{a}$  vary we obtain the twelve contour plots in fig. \ref{fig7}. The 
first six plots predict energy resolution while the last six plots refer to 
the detectors time constant. Each column of contour plots belongs to one of 
the three different models.

\section{Conclusions}
To improve the performance of microcalorimeters and bolometers it is
important to accurately understand how this depends on the fabrications
parameters. Significant improvements in detectors performance have, in 
fact, been achieved by optimizing the design based on an accurate model of 
the detector \cite{G1}. In this paper we derived detailed theoretical models to
describe the behavior of two different detector architectures. The use 
of block diagram algebra has allowed us to present the results in an 
analytical form that can be easily and immediately utilized by investigators to
 improve the design of their detectors. Furthermore, the contour plots in 
fig.~\ref{fig7} provide an example of how the equations derived in this paper 
can be utilized to predict the characteristics of detectors.

\newpage
\section*{Tables}
Table 1: Values of the fixed parameters needed to produce the contour
 plots of fig.~\ref{fig7}
\section*{Figures}
FIG. 1: Thermal architecture of Model 1.

FIG. 2: Typical readout circuit. Notice that if $R_{L}<<R$ the detector 
is voltage biased, if $R_{L}>>R$ the detector is current biased.

FIG. 3: Block diagram representing Model 1.

FIG. 4: Block diagram including noise contributions 
for Model 1.

FIG. 5: Thermal architecture of Model 2.

FIG. 6: Block diagram including noise contributions 
for Model 2.

FIG. 7: Contour plots of how the energy resolution and time 
constant of each model change with respect to the heat capacity $C_{a}$ 
and the thermal conductivities $G$ and $G_{a}$. These plots were constructed 
using the fixed parameters in Table 1.

\begin{figure}
\begin{center}
\includegraphics[width=1\textwidth]{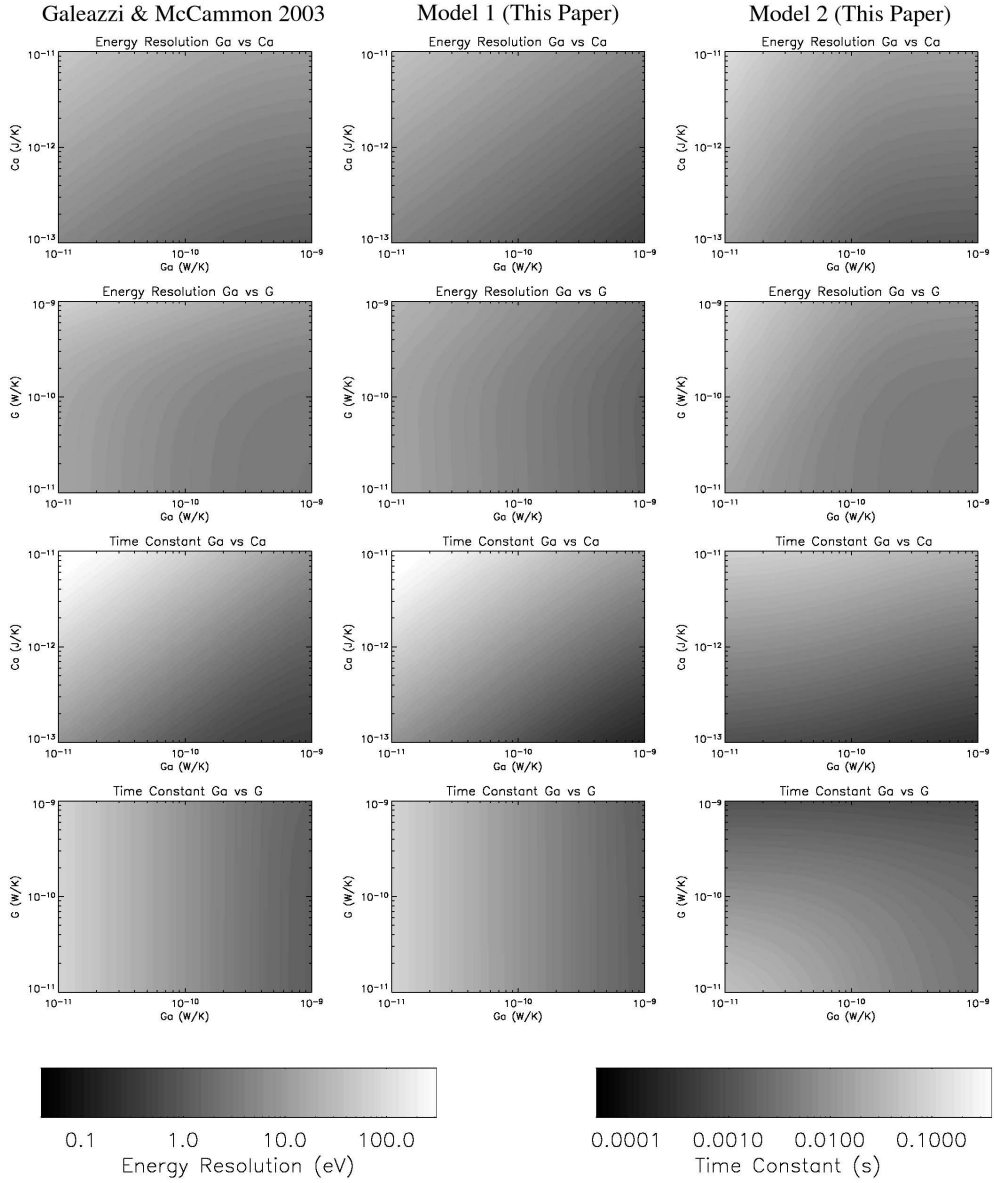}
\end{center}
\caption{\footnotesize Contour plots of how the energy resolution and time 
constant of each model change with respect to the heat capacity $C_{a}$ 
and the thermal conductivities $G$ and $G_{a}$. These plots were constructed 
using the fixed parameters in Table 1.  }
\label{fig7}
\end{figure}

\newpage
\section*{Table 1: Values of the fixed parameters needed to produce 
the contour plots in fig \ref{fig7}}
\begin{tabular}{|c|c|}
\hline
 Parameter & Value \\
\hline
$R$ & $5 m\Omega$ \\
\hline
$R_{L}$ & $0.2 m\Omega$\\
\hline
$V_{bias}$ & $1.0848 \times  10^{-7} V$\\
\hline
$\alpha$ & $100$\\
\hline
$C_{a}$ & $1 pJ/K$ \\
\hline
$C_{l}$ & $4.911 \times  10^{-5} pJ/K$\\
\hline
$C_{e}$ & $0.154 pJ/K$\\
\hline
$T_{s}$ & $0.1 K$\\
\hline
$G_{e-l}(T_{l})$ & $5 \times 10^{-10} W/K$\\
\hline
$G_{e-l}(T_{e})$ & $5.87 \times 10^{-10} W/K$\\
\hline
$G$ & $1 \times 10^{-10} W/K$\\
\hline
\end{tabular}
\newpage

\newpage

\newpage

\newpage

\newpage

\newpage

\newpage

\end{document}